\documentclass[aps,prd,twocolumn,groupedaddress,showpacs]{revtex4}
\usepackage{graphicx}

\begin{document}


\title{Dark Matter in the Solar System}


\author{X. Xu}
\email[]{xxu@as.arizona.edu}
\author{E. R. Siegel}
\email[]{ethan@gravity.as.arizona.edu}
\homepage[]{http://startswithabang.com/}
\affiliation{Steward Observatory, University of Arizona, 933 N. Cherry Ave., Tucson, AZ 85721}


\date{\today}

\begin{abstract}
We determine the density and mass distribution of dark matter within our
Solar System.  We explore the three-body interactions between dark matter
particles, the Sun, and the planets to compute the amount of dark matter
gravitationally captured over the lifetime of the Solar System.  We provide
an analytical framework for performing these calculations and detail our
numerical simulations accordingly.  We find that the local density of dark
matter is enhanced by between three and five orders of magnitude over
the background halo density, dependent on the radial distance from the Sun.
This has profound implications for terrestrial direct dark matter detection 
searches.  We also discuss our results in the context of gravitational
signatures, including existing constraints, and find that dark matter 
captured in this fashion is not responsible for the Pioneer anomaly.
We conclude that dark matter appears to, overall, play a much more important 
role in our Solar System than previously thought.
\end{abstract}

\pacs{95.30.Sf, 95.35.+d, 96.50.Pw, 98.35.Gi, 98.80.-k}

\maketitle


\section{\label{Sec1}Introduction}

Accounting for about $23\%$ of the energy density of the Universe 
\cite{Komatsu:2008hk}, 
dark matter is an integral part of our surroundings. It dominates the 
more familiar, baryonic components, which comprises only  $4.4\%$ of the 
Universe, on the largest scales.  
Achieving an understanding of this perplexing dark element is one of
the most compelling unsolved problems in modern astrophysics.

The astrophysical evidence for the existence of dark matter is overwhelming,
as observations of the cosmic microwave background \cite{Komatsu:2008hk}, 
the power spectrum of the Universe \cite{2dFSDSS}, and colliding galaxy
clusters \cite{Clowe:2006} all point towards the same conclusion.  With an
understanding that a dark, pressureless, fluid-like component permeates
the Universe, astrophysicists have successfully simulated the large-scale
processes of structure formation in the context of a $\Lambda$CDM Universe
\cite{Monstrosity}.  More recently, some attention has been given to the
dark matter substructure formed on subgalactic scales, down to scales of
order $10^{-2} \,$pc \cite{lilCDM}.  However, relatively little consideration
has been given to the distribution of dark matter within our own Solar System.

Yet, dark matter may prove to be profoundly important in our Solar System
for both its additional gravitational effects on planets and other orbiting
bodies \cite{Monstrosity2,Gron:1996,Frere:2008} as well as 
the motions 
of spacecraft \cite{Anderson:1998,Adler:2008}.  Furthermore, a knowledge of 
the density and velocity of dark matter particles is particularly important
for terrestrial direct detection experiments \cite{Aprile:2005ww}.

In this paper, we model the Solar System and the dark matter that it
encounters in order to quantify how much dark matter we expect the Solar
System to have captured over its lifetime.  Through favorable three-body
gravitational interactions between a dark matter particle, the Sun, and
any of the planets, a non-zero and possibly significant fraction of the
dark matter passing through the Solar System will become gravitationally
bound to it.  The remainder of this paper is focused on solving this problem,
and is laid out as follows: section 2 details the model chosen for the Solar 
System and galactic properties, and provides analytic details of our 
calculations.  Section 3 sets forth the computations undertaken to 
successfully determine the probability of binding dark matter to the Solar
System.  Section 4 presents our results, including the density and mass
distributions of dark matter with respect to distance from the Sun.  
We recommend that anyone not interested in the details of our calculations
skip directly to section 4.  Finally,
section 5 concludes with a discussion of the conclusions reached from our
analysis, detailing significant implications for dark matter detection and
presenting a comparison of our results with the current experimental 
and observational constraints.

\section{\label{Sec2}Dark Matter Capture}

Our Sun (and hence our Solar System) is presently moving through the galaxy
with well-known parameters \cite{Kerr:1986} that have changed little, despite 
refinements in measurements \cite{Olling:1998}, over many years.  
More recently, we have been able to determine that our Milky Way, 
like all comparable galaxies, is pervaded by a dark matter halo 
with a specific density profile \cite{PROF}.  $N$-body simulations also
provide insights into modeling the Milky Way \cite{Klypin:2002}.
Based on the fact that our Sun is not an isolated body, but rather has eight
planets as well as a number of other, less significant bodies gravitationally
bound to it, a non-zero fraction of this dark matter will be captured by
favorable three-body interactions between a dark matter particle, the Sun, and
an orbiting body.

In order to compute the amount of dark matter captured by the Solar System
over the course of its $4.5 \, \mathrm{Gyr}$ history, we assume the galactic
parameters shown in Table \ref{MilkyWay} below.  
\begin{table}[h]
\begin{center}
\begin{tabular}{|l|c|}
\hline
Parameter & Value \\
\hline
\hline
Oort's Constant $A$ $(\mathrm{km} \, \mathrm{s}^{-1} \, \mathrm{kpc}^{-1})$ 
& 14.4 \\
\hline
Solar Radius $r_\odot$ $(\mathrm{kpc})$ & 8.0 \\
\hline
Speed of Local Standard of Rest (LSR) $v_\mathrm{lsr}$ 
& 200 \\
\hline
Sun's speed relative to LSR $v_{\odot \mathrm{,lsr}}$ 
& 20 \\
\hline
Velocity relative to the Galactic Plane 
$v_{z}$ 
& 7 \\
\hline
Solar Period relative to the Galactic Plane $t_\mathrm{osc}$ $(\mathrm{Myr})$ 
& 63 \\
\hline
Local Dark Matter density $\rho_\mathrm{DM} (r_\odot)$ 
$(M_\odot \, \mathrm{pc}^{-3})$ & $ 0.009 $ \\
\hline
Mass within $r_\odot$ of Galactic center $M_\mathrm{enc}(r_\odot)$ 
$(10^{10} \, M_\odot)$ & $ 9 $ \\
\hline
Local $rms-$velocity of Dark Matter $v_\mathrm{rms}(r_\odot)$ 
& 220 \\
\hline
\end{tabular}
\caption{\label{MilkyWay}
The local galactic and dark matter parameters used for the Milky Way and the
present values of the Sun's distance and velocity components.  
Unless otherwise noted, all velocities are in units of 
$\mathrm{km} \, \mathrm{s}^{-1}$.}
\end{center}
\end{table}
We then find that the Sun 
moves through the galaxy with a velocity $(v_\odot)$ with components given
in cylindrical coordinates:
\begin{eqnarray}
\label{Vsun}
&v_{\odot \mathrm{,} \rho}&=A r_\odot \cos{(2 l)} \mathrm{,} \nonumber\\
&v_{\odot \mathrm{,}  \phi}&=v_\mathrm{lsr} - v_{\odot \mathrm{,lsr}} 
\sin{(2 l)} \mathrm{,} \nonumber\\
&v_{\odot \mathrm{,}  z}&= v_z \sin{\left( \frac{2 \pi t}{t_\mathrm{osc}} 
\right) } \mathrm{,}
\end{eqnarray}
where $l$ is the galactic longitude, $t$ is a time coordinate, and 
the remaining parameters are defined in Table \ref{MilkyWay}.
The individual dark matter particles are assumed to follow a 
Maxwell-Boltzmann distribution \cite{Alenazi:2006} with a 
probability distribution function $f(v)$ given by
\begin{equation}
\label{MaxBoltz}
f(v)=\sqrt{\frac{54}{\pi}} \frac{v^2}{v^3_\mathrm{rms}(r)} 
e^{ - \frac{3}{2} \frac{v^2}{v^2_\mathrm{rms}(r)}} \mathrm{,}
\end{equation}
where the local $rms$-velocity $v_\mathrm{rms}(r_\odot)$ is given in 
Table \ref{MilkyWay}.

Therefore, the Sun sweeps out a predictable three-dimensional path
over its $4.5 \, \mathrm{Gyr}$ history.  With the dark matter having a
local density $\rho_\mathrm{DM}(r_\odot)$, a $rms$-velocity 
$v_\mathrm{rms}(r_\odot)$ and a velocity distribution as given above, the
fraction of dark matter captured can be calculated in a straightforward
fashion via modeling of the Solar System and the dark matter particles passing
through it.  However, the number of dark matter particles
encountered is far too large and the rate of particle capture is far
too small to effectively simulate using $N$-body
methods.  We are therefore forced to use analytic
approximations to shape this problem into a tractable one.

We begin by considering an ensemble of dark matter particles at infinity 
each with a speed $|v_\mathrm{DM}|$ chosen from the Maxwell-Boltzmann
distribution and a random orientation $\hat{v}_\mathrm{DM}$. We also consider
the Sun, and a single planet with 
mass $m_p$, distance from the Sun $r_p$, and a circularized velocity 
around the Sun $v_p$.  (This analysis will be repeated eight times, 
once for each of the planets.)  We first choose a very large $r_\infty$
to be the vector distance from the dark matter particle to the Sun,
so that $|r_\infty| \gg |r_p|$, but small enough that when we compute the
dark matter particle's angular momentum with respect to the Sun, we 
get a reasonable (i.e., non-infinite) value.  

We then perform a coordinate transformation to shift into the Sun's rest 
frame, computing the velocity of the dark matter at infinity $(v_\infty)$ 
in that frame:
\begin{eqnarray}
\label{Vinf}
v_{\infty \parallel} &=& \frac{(v_\mathrm{DM} - v_\odot) \cdot r_\infty}{|r_\infty |} \nonumber\\
v_{\infty \perp} &=& \frac{(v_\mathrm{DM} - v_\odot) \times r_\infty}{|r_\infty |} \mathrm{,}
\end{eqnarray}
where $v_{\infty \parallel}$ and $v_{\infty \perp}$ are the components
of the dark matter particle's velocity parallel and perpendicular to the
Sun's, respectively.

The dark matter particles we are interested in, as far as the 
possibility of gravitational capture goes, are those that pass within 
a distance $r_p$ of the Sun.  The dark matter particles, at infinity,
will have an angular momentum $L$ given by
\begin{equation}
\label{L}
L = m v_{\infty \perp} r_\infty \mathrm{,}
\end{equation}
where $m$ is the mass of a dark matter particle.  We note that this mass
is completely unimportant in our analyses, as it does not enter into
any of our equations; only the combination $L/m$ appears.  These particles are 
all in hyperbolic orbits around the Sun initially, with 
eccentricities $\epsilon$ given by
\begin{equation}
\label{ecc}
\epsilon = \sqrt{1 + \frac{v_\infty^2 ( L/m )^2}{G^2 M_\odot^2}} \mathrm{.}
\end{equation}
We then find that any particle that meets the following condition will
pass within a distance $r_p$ of the Sun:
\begin{equation}
\label{cond1}
\epsilon \geq \left| \frac{ (L/m)^2 }{G M_\odot r_p} - 1 \right| \mathrm{.}
\end{equation}
Upon encountering the planet, the particle may receive boosts (or 
anti-boosts) at two points during the interaction. The first occurs at entry
into the sphere of radius $r_p$ and the second occurs upon exit of the sphere.
Taking both of these opportunities
into account, we consider the approximation that the planet's 
position at any time is given by a random location on the sphere of radius
$r_p$.  We determine that of the particles that pass through the sphere 
of radius $r_p$, a fraction $r_b^2 / r_p^2$ of those will gravitationally
encounter the planet, where $r_b$ equals
\begin{equation}
\label{r_b}
r_b \equiv 1.15 r_p \left( \frac{m_p}{M_\odot} \right)^{1/3}
\mathrm{,}
\end{equation}
with $r_b$ defined to be the radius of a ``sphere of influence'' of
a smaller gravitational body relative to a larger one \cite{Kislik}.

\begin{figure}
\includegraphics[width=\columnwidth]{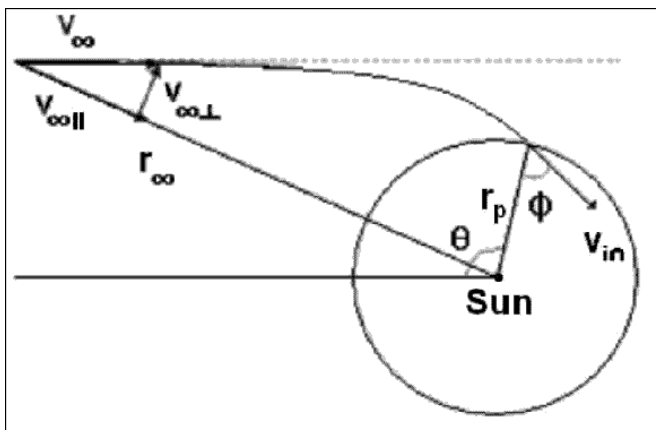}%
\caption{
Geometric setup for a dark matter particle approaching the Solar System 
from infinity in the Sun's rest frame.  The semi-major axis of the planet 
in question's orbit is shown by $r_p$, with the velocity that the 
particle strikes the 
imaginary sphere of radius $r_p$ given by $v_\mathrm{in}$.  Through the
conservation of angular momentum, the angle $\theta$ can be determined, as
shown in equation (\ref{rdeflection}), and the angle that the velocity
$v_\mathrm{in}$ makes with $r_p$, defined as $\phi$, is given by equation
(\ref{vdeflection}).}
\label{geometry}
\end{figure}

We then need to determine whether the gravitational encounter is favorable
enough to transition the dark matter particle from an unbound, hyperbolic
orbit around the Sun to a bound, elliptical one.  We begin by determining
the velocity of the incoming dark matter particle at a distance $r_p$ from
the Sun $(\equiv v_\mathrm{in})$.  The geometry of an incoming dark matter
particle as it enters the Solar System, possibly encountering a given planet's
sphere of influence, is illustrated in Figure \ref{geometry}.
This velocity will have a magnitude given by the conservation of energy,
\begin{equation}
\label{vin}
|v_\mathrm{in}| = \sqrt{v_\infty^2 + \frac{2 G M_\odot}{r_p}} \mathrm{,}
\end{equation}
and we are interested in determining the components parallel to the radial
vector towards the Sun $(v_{\mathrm{in}\parallel})$ and perpendicular to it
$(v_{\mathrm{in}\perp})$.  We can determine these, in the Sun's rest frame,
by first finding the angle of deflection $(\theta)$ caused by the 
gravitational force on the position of the dark matter particle's azimuthal
angle from infinity to $r_p$, 
\begin{equation}
\label{rdeflection}
\theta = -\cos^{-1} \left( \frac{ \frac{(L/m)^2}{G M_\odot r_p} - 1}{\epsilon}
\right) + \cos^{-1} \left( - \frac{1}{\epsilon} \right) \mathrm{,}
\end{equation}
and then, through the conservation of angular momentum, the angle of
deflection $(\phi)$ of the particle's velocity,
\begin{equation}
\label{vdeflection}
\phi = \sin^{-1} \left[ \left| \frac{L/m}{v_\mathrm{in} r_p} \right| 
\left( \sin (\theta) \pm \cos (\theta) \sqrt{ 
\frac{ v_\mathrm{in}^2 r_p^2 }{ (L/m)^2 } - 1} \right) \right]
\mathrm{,}
\end{equation}
where $\phi$ is chosen to be the smaller in magnitude of the two possible 
angles.  From equations (\ref{vin}-\ref{vdeflection}), we can then determine
the components of $v_\mathrm{in}$ to be
\begin{eqnarray}
\label{vincomponents}
v_{\mathrm{in}\parallel} &=& |v_\mathrm{in} \left[ \sin (\theta) \cos (\phi) 
- \cos (\theta) \sin (\phi) \right] | \mathrm{,} \\
v_{\mathrm{in}\perp} &=& \sqrt{ v_\mathrm{in}^2 - v_{\mathrm{in}\parallel}^2 }
\mathrm{.}
\end{eqnarray}

Assuming that the particle does get within a distance $r_b$ of the planet
(which, as stated above, it does with probability $r_b^2 / r_p^2$), we then
need to determine the gravitational effect of the planet on the dark
matter particle's orbit with respect to the Sun.  We assume that the planet
is positioned randomly in space at a distance $r_p$ from the Sun, and moves 
with a speed $|v_p|$ in a random direction perpendicular 
to $v_{\mathrm{in}\parallel}$.  

We then switch to the planet's rest frame, and obtain, for the dark matter
particle, a velocity $(v_\mathrm{dm})$ with components
\begin{eqnarray}
\label{vdmplanet}
v_{\mathrm{dm,}x} &=& v_{\mathrm{in}\parallel} \mathrm{,} \\
v_{\mathrm{dm,}y} &=& v_{\mathrm{in}\perp} - v_p \sin (\alpha) \mathrm{,} \\
v_{\mathrm{dm,}z} &=& -v_p \cos (\alpha)
\mathrm{,}
\end{eqnarray}
where $\alpha$ is a random angle between $0$ and $2 \pi$, and a position
with respect to the planet $(r_\mathrm{dm})$ with components
\begin{equation}
\label{rdmplanet}
r_\mathrm{dm} = \langle \sqrt{ r_b^2 - r_{\mathrm{dm,}y}^2 - 
r_{\mathrm{dm,}z}^2 } \mathrm{,} \, r_{\mathrm{dm,}y} \mathrm{,} \,
r_{\mathrm{dm,}z} \rangle
\mathrm{,}
\end{equation}
where $r_{\mathrm{dm,}y}$ and $r_{\mathrm{dm,}z}$ are randomly chosen to
lie within a circle of radius $r_b$ in the $yz-$plane.  We also note that
in this rest frame, the Sun appears to move with a velocity
\begin{equation}
\label{vsunplanet}
v_\odot = \langle 0 \mathrm{,} \, -v_p \sin (\alpha) \mathrm{,} \,
-v_p \cos (\alpha) \rangle \mathrm{.}
\end{equation}

We then compute the components of $v_\mathrm{dm}$ parallel and perpendicular
to $r_\mathrm{dm}$,
\begin{eqnarray}
\label{vdmandrdm}
v_{\mathrm{dm}\parallel} &=& \left| \frac{v_\mathrm{dm} \cdot r_\mathrm{dm}}
{r_b} \right| \mathrm{,} \\
v_{\mathrm{dm}\perp} &=& \left| \frac{v_\mathrm{dm} \times r_\mathrm{dm}}
{r_b} \right| \mathrm{,}
\end{eqnarray}
and the components of $v_\odot$ with respect to these new directions,
\begin{eqnarray}
\label{vsunandvdm}
v_{\odot \parallel} &=& \frac{v_\odot \cdot r_\mathrm{dm}}{|r_b|} \mathrm{,} \\
v_{\odot \perp} &=& v_\odot \cdot \left( \frac{v_\mathrm{dm} \times 
r_\mathrm{dm}}{ | v_{\mathrm{dm}\perp} \, r_b | } \right) \mathrm{,} \\
v_{\odot \mathrm{3d}} &=& \sqrt{v_\odot^2 - v_{\odot \parallel}^2 
- v_{\odot \perp}^2 }
\mathrm{,}
\end{eqnarray}
where $v_{\odot \mathrm{3d}}$ is the component of $v_\odot$ orthogonal to
both $\hat{v}_{\mathrm{dm}\parallel}$ and $\hat{v}_{\mathrm{dm}\perp}$.
In this coordinate system, it is easy to compute the angle $(\beta)$ that
the planet causes the dark matter particle to deflect by,
\begin{equation}
\label{beta}
\beta = 2 \tan^{-1} \left| \frac{G \, m_p}{ r_b \, v_{\mathrm{dm}\perp} \, 
v_\mathrm{dm} } \right| \mathrm{,}
\end{equation}
where the final outgoing velocity of the dark matter particle 
$(v_\mathrm{out})$ is then given by
\begin{eqnarray}
\label{vout}
v_\mathrm{out} &=& \langle v_{\mathrm{dm}\parallel} \cos (\beta) + 
v_{\mathrm{dm}\perp} \sin (\beta) \mathrm{,} \nonumber\\
&& - v_{\mathrm{dm}\parallel} 
\sin (\beta) + v_{\mathrm{dm}\perp} \cos (\beta) \mathrm{,} \, 0 
\rangle \mathrm{.}
\end{eqnarray}
At last, we can compute the total final speed of the dark matter
particle $(|v_f|)$, as it leaves the sphere of influence of the planet, 
in the rest frame of the Sun,
\begin{equation}
\label{vf}
|v_f| = \sqrt{ (v_\mathrm{out} - v_\odot) \cdot (v_\mathrm{out} - v_\odot) }
\mathrm{.}
\end{equation}
The particle will be gravitationally captured if
\begin{equation}
\label{capcondition}
v_f < v_\mathrm{escape}(r_p) \equiv \sqrt{ \frac{2 G M_\odot}{r_p} }
\mathrm{,}
\end{equation}
and become bound in an elliptical orbit about the Sun with semi-major 
axis
\begin{equation}
\label{semi-majoraxis}
a = \left( \frac{|v_f|^2}{G M_\odot} - \frac{2}{r_p} \right)^{-1}\mathrm{.}
\end{equation}

A point to note is that we do not consider any further
three-body interactions between captured dark matter, the planets and the Sun.
Once the dark matter is captured, we assume it remains captured without
any further gravitational interactions of significance.
However, such a back-reaction will exist, and could potentially decrease 
the amount of dark matter bound to the Solar System by a significant amount.

\section{\label{Sec3}Computations}

For each of the eight planets in our Solar System, we perform the 
calculations outlined in section \ref{Sec2}.  The planets are assumed 
to have the parameters shown below in Table \ref{8planets}.
\begin{table}[h]
\begin{center}
\begin{tabular}{|l|c|c|c|}
\hline
Planet & Distance to Sun & Mass & Speed \\
\hline
\hline
Mercury & $5.79 \times 10^7$ km & $1.68 \times 10^{-7} \, M_\odot$ & $48 \, \mathrm{km} \, 
\mathrm{s}^{-1}$ \\
\hline
Venus & $1.08 \times 10^8$ km & $2.46 \times 10^{-6} \, M_\odot$ & $35 \, \mathrm{km} \, 
\mathrm{s}^{-1}$ \\
\hline
Earth & $1.496 \times 10^8$ km & $2.99 \times 10^{-6} \, M_\odot$ & $30 \, \mathrm{km} \, 
\mathrm{s}^{-1}$ \\
\hline
Mars & $2.28 \times 10^8$ km & $3.21 \times 10^{-7} \, M_\odot$ & $24 \, \mathrm{km} \, 
\mathrm{s}^{-1}$ \\
\hline
Jupiter & $7.78 \times 10^8$ km & $9.50 \times 10^{-4} \, M_\odot$ & $13 \, \mathrm{km} \, 
\mathrm{s}^{-1}$ \\
\hline
Saturn & $1.43 \times 10^9$ km & $2.86 \times 10^{-4} \, M_\odot$ & $9.6 \, \mathrm{km} \, 
\mathrm{s}^{-1}$ \\
\hline
Uranus & $2.87 \times 10^9$ km & $4.40 \times 10^{-5} \, M_\odot$ & $6.8 \, \mathrm{km} \, 
\mathrm{s}^{-1}$ \\
\hline
Neptune & $4.50 \times 10^9$ km & $5.11 \times 10^{-5} \, M_\odot$ & $5.4 \, \mathrm{km} \, 
\mathrm{s}^{-1}$ \\
\hline
\end{tabular}
\caption{\label{8planets}
Planet-Sun distances, planetary masses and speeds for the eight Solar 
System planets considered in this analysis.  Distances are given in 
units of km, masses are given in units of solar masses, where 
$M_\odot = 1.9884 \times 10^{30} \, \mathrm{kg}$, and speeds in units of 
$\mathrm{km} \, \mathrm{s}^{-1}$.}
\end{center}
\end{table}
We begin by creating a cumulative distribution function for the 
Maxwell-Boltzmann distribution (equation (\ref{MaxBoltz})), obtaining
cumulative probability $P(v)$ as a function of the dark matter's velocity,
\begin{eqnarray}
\label{cdf}
P(v) &=& \int_0^v f(v ' ) dv ' \nonumber\\
&=& \mathrm{erf} \left( 
\sqrt{\frac32} \frac{v}{v_\mathrm{rms}} \right) - \sqrt{ \frac{6}{\pi} }
\frac{v}{v_\mathrm{rms}} e^{ - \frac{3}{2} \frac{v^2}{v^2_\mathrm{rms}(r)}}
\mathrm{,}
\end{eqnarray}
where $P(v)$ is the probability of finding a dark matter particle with
velocity less than or equal to $v$.  From this cumulative distribution
function, we determine the velocity with respect to the Sun and keep only
those dark matter particles that meet the condition
\begin{equation}
\label{condition1}
|v_\infty| \leq 2 \left( v_p^2 + v_p \sqrt{\frac{2 G M_\odot}{r_p}} 
\right)^{1/2}
\mathrm{,}
\end{equation}
where $v_p$ and $r_p$ are the values for the appropriate planet as given in
Table \ref{8planets}, and $v_\infty$ is the dark matter particle's speed with 
respect to the Sun's reference frame at infinity.  We choose the
condition in equation (\ref{condition1}) because even the most
favorable gravitational interaction with the planet can only decrease
the speed of an incoming dark matter particle by $2v_p$.  The condition in
equation (\ref{condition1}) is such that a particle reaching a distance
$r_p$ from the Sun will have a velocity no greater than $2v_p$.  We generate 
at least one million unique particles that satisfy this condition for each 
planet.

Keeping track of the total number of particles simulated before generating
the one million we seek, we compute the probability of particles satisfying
the preceding speed constraint. We then generate
random directions for a particle, with an initial position 
at $r_\infty$, and a speed chosen randomly from the the set of one 
million particles.  We simulate a total of 100 billion particles for each
of the Jovian planets and from 300 to 600 billion particles for each of the 
inner, rocky planets, which require more due to their much smaller 
distances from the Sun.  We adopt 
$r_\infty \equiv 2.67 \times 10^{15} \, \mathrm{m}$, placing it orders of
magnitude beyond the orbit of Neptune, but still close enough so that
a reasonable number of simulated particles will interact with each planet,
based on the condition in equation (\ref{cond1}).  We then compute the
location and velocity of all of the particles that do pass within a distance
$r_p$ of the Sun (correcting for the fact that the $r_\infty$ chosen is
not actually infinite), as well as the probability of randomly
generated particles passing within $r_p$ of the Sun.

We then reduce that probability further by a factor of $r_b^2 / r_p^2$, as 
discussed in section \ref{Sec2}, as only that fraction of the dark matter 
particles passing within a distance $r_p$ of the Sun will pass within a
distance $r_b$ of the planet in question.

We then generate, from the particles that have passed all the cuts up until
now, a random set of positions for the planet within a three-dimensional
distance $r_b$ of the dark matter particle, according to equation 
(\ref{rdmplanet}), along with velocity directions for the planet in accordance
with equations (\ref{vdmplanet}-15).  From the particles that survive the
earlier cuts, we choose enough random positions and velocity directions to
generate one billion particles for this final step.  By boosting to the 
planet's rest frame, calculating the change in direction due to the 
gravitational interaction between the dark matter particle and the 
planet, and then boosting back to the Sun's rest frame, we obtain the 
final speed of the particle.  We tabulate the particles that
become bound to the Sun as a result of this final interaction, and 
compute both the total probability of gravitational capture and the
distribution of the semi-major axes of the captured particles.

\section{\label{Sec4}Results}

Given the $r_\infty$ chosen in section \ref{Sec3} and the fact that the
Solar System has had approximately 4.5 billion years (defined as the lifetime
of the Solar System, $t_\mathrm{SS}$) to accrue dark matter
via this mechanism, we determine that the total amount of dark matter
encountered is given by
\begin{eqnarray}
\label{totalDM}
M_\mathrm{DM} &=& \rho_\mathrm{DM}(r_\odot) \, V \nonumber\\
&=& \rho_\mathrm{DM}(r_\odot) \pi r_\infty^2 \bar{v}_\odot t_\mathrm{SS}
\simeq 203 \, M_\odot
\mathrm{,}
\end{eqnarray}
where $\bar{v}_\odot$ is the average velocity of the Sun with respect to the
galaxy, determined to be $208 \, \mathrm{km} \, \mathrm{s}^{-1}$ using
a time average of the data from equation (\ref{Vsun}) and Table \ref{MilkyWay}.
\begin{table}[h]
\begin{center}
\begin{tabular}{|c|c|c|}
\hline
Planet & Fraction Captured & Mass Captured \\
\hline
\hline
Mercury&$1.03 \times 10^{-16}$ & $2.09 \times 10^{-14} \, M_\odot$\\
\hline
Venus&$8.71 \times 10^{-16}$ & $1.77 \times 10^{-13} \, M_\odot$\\
\hline
Earth&$9.41 \times 10^{-16}$ & $1.91 \times 10^{-13} \, M_\odot$\\
\hline
Mars&$2.91 \times 10^{-16}$ & $5.91 \times 10^{-14} \, M_\odot$\\
\hline
Jupiter&$1.23 \times 10^{-13}$ & $2.50 \times 10^{-11} \, M_\odot$\\
\hline
Saturn&$7.06 \times 10^{-14}$ & $1.43 \times 10^{-11} \, M_\odot$\\
\hline
Uranus&$2.87 \times 10^{-14}$ & $5.83 \times 10^{-12} \, M_\odot$\\
\hline
Neptune&$3.98 \times 10^{-14}$ & $8.08 \times 10^{-12} \, M_\odot$\\
\hline
\end{tabular}
\caption{\label{Results}
Fraction and total mass of all dark matter particles captured due to 
gravitational interactions with each of the eight planets.
Although the absolute numbers are much smaller for the inner, rocky planets
compared to the Jovians, they are still significant for determining 
the densities of dark matter, as they dominate at radii smaller than
half of Jupiter's semi-major axis.}
\end{center}
\end{table}
Our calculations and computations then allow us to determine what fraction 
of this total mass winds up getting gravitationally captured via these
three-body interactions.  Our results are presented in Table \ref{Results}.

The two most important factors contributing to capture are the mass and 
the orbital semi-major axis of the planet. The more massive and the further 
away a planet is from the Sun, the more effective it will be at capturing dark 
matter, as the cross-section for favorable gravitational interactions is 
given by the radius of the sphere of influence $(r_b)$ from equation
(\ref{r_b}). 

\begin{figure}
\includegraphics[width=\columnwidth]{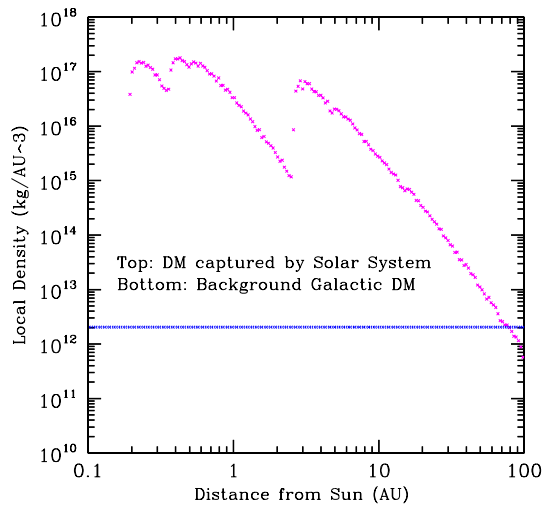}%
\caption{Dark matter density vs$\mathrm{.}$ distance from the Sun in our Solar
System at the present day.  Density is presented in units of $\mathrm{kg} \,
\mathrm{AU}^{-3}$, with distances given in units of $\mathrm{AU}$.}
\label{denplot}
\end{figure}

Equation (\ref{semi-majoraxis}) allows us to obtain a measure of the spatial
distribution of the captured dark matter particles.  In Figure \ref{denplot},
we illustrate the local
density of dark matter, taking the semi-major axis of the captured dark
matter particle as a proxy for its position, as a function of distance
from the Sun.  For comparison, the background galactic halo density is
also shown.  Of particular note is the dark matter density at 
$1 \, \mathrm{AU}$, which is greater than the background dark matter density
(from the underlying galactic halo) by a factor of $1.63 \times 10^{4}$.

\begin{figure}
\includegraphics[width=\columnwidth]{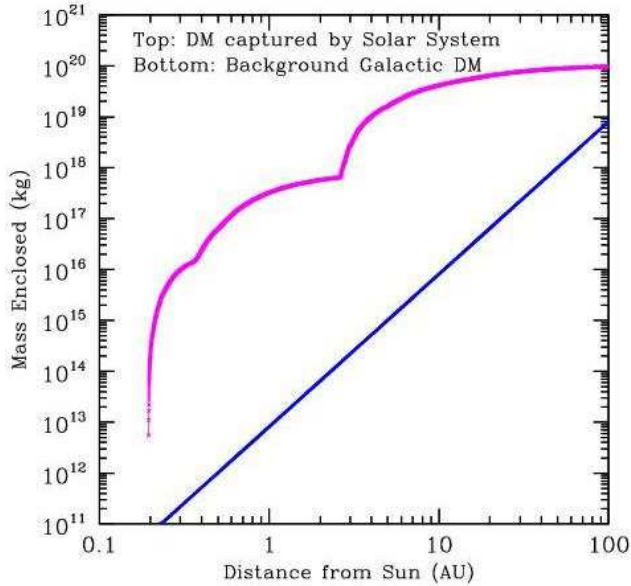}%
\caption{Cumulatively enclosed mass as a function of distance from the
Sun at present.  This is representative of the additional mass gravitationally
felt by an object in orbit around the Sun due to the presence of captured
dark matter.}
\label{massplot}
\end{figure}

Figure \ref{massplot} indicates the total mass of dark matter enclosed 
within a certain radius from the Sun due to both the captured dark matter
and the underlying galactic halo. Between approximately $0.2 \, \mathrm{AU}$
and $100 \, \mathrm{AU}$, the amount of dark matter bound
to the Solar System is much more massive and dense than the background halo.
Within the orbit of Mercury, Earth, Mars, and Neptune, we find that 
approximately $1.91 \times 10^{16} \, \mathrm{kg}$, $3.23 \times 10^{17}
 \, \mathrm{kg}$, $4.87 \times 10^{17} \, \mathrm{kg}$, and $7.69 \times
10^{19} \, \mathrm{kg}$ of dark matter is enclosed, respectively.  
The total amount of matter dark bound to the Solar
System as the result of gravitational capture is $1.07 \times 10^{20} \,
\mathrm{kg}$, or $1.78 \times 10^{-5}$ Earth masses.

\section{\label{Sec5}Discussion}

We predict the presence of a new component of dark matter within the Solar
System due to gravitational capture.
We find that, within the orbit of Neptune, $7.69 \times 10^{19} \, 
\mathrm{kg}$ of dark matter has become bound
to our Solar System due to the capture mechanism over its lifetime. 
This is about a factor of $300$ greater than the background mass from the 
galactic halo, as shown in Figure \ref{massplot}.  The density of the
captured dark matter is also significantly enhanced in comparison with the 
halo, as demonstrated in Figure \ref{denplot}.  At the Earth's orbital radius,
density is enhanced by more than four orders of magnitude over the local
halo density, with a value of $3.3 \times 10^{16} \, \mathrm{kg} \, 
\mathrm{AU}^{-3}$. Such elevated levels of dark matter have important
implications for direct detection experiments and can be tested as a 
potential explanation for spacecraft anomalies.

Direct detection searches for dark matter rely predominantly on nuclear
recoils \cite{Aprile:2005ww}, where the rate of interaction is dependent
on the dark matter's density, velocity, and interaction cross-section 
(which itself may have a velocity dependence).  
Our determination of the local dark matter density and velocity distribution
are of great importance to direct detection experiments. 
The most recent calculations that have been carried out assume that the 
properties of dark matter at the Sun's location are derived directly from the
galactic halo \cite{Kamionkowski:2008}.  By comparison, we find that
terrestrial experiments should also consider a component of dark matter
with a density $16\mathrm{,}000$ times greater than the background halo
density, albeit with a $v_\mathrm{rms}$ on the order of the Earth's
orbital speed $(30 \, \mathrm{km} \,
\mathrm{s}^{-1})$, about a factor of seven smaller than the $rms$-velocity of
the unbound halo particles.  If this dark matter is not efficiently ejected
by future interactions after the initial capture, the gravitationally bound
component of dark matter may wind up dominating the signal expected from 
future direct detection experiments. 

One method that has been used to constrain the amount of dark matter in
the Solar System has been careful, long-term observations of planetary
and satellite motions.  Constraints have been obtained both from planetary
orbital data 
\cite{Monstrosity2,Frere:2008} and
perihelion shift observations 
\cite{Gron:1996,Khriplovich:2007}. The most stringent results 
for the dark matter density near Earth constrain $\rho_\mathrm{DM}$ to
be less than $6.0 \times 10^{16} \, \mathrm{kg} \, \mathrm{AU}^{-3}$ from 
orbital data \cite{Frere:2008}.  For the densities near Mercury and Mars,
perihelion precession provides the tightest constraints, yielding upper
limits on $\rho_\mathrm{DM}$ of $8.7 \times 10^{18} \, \mathrm{kg} \, 
\mathrm{AU}^{-3}$ near Mercury and $5.4 \times 10^{16} \, \mathrm{kg} \, 
\mathrm{AU}^{-3}$ near Mars \cite{Khriplovich:2007}.  Our results satisfy
these constraints, as we predict the densities near Mercury, Earth, and Mars
to be $1.5 \times 10^{17} \, \mathrm{kg} \, \mathrm{AU}^{-3}$, $3.3 \times 
10^{16} \, \mathrm{kg} \, \mathrm{AU}^{-3}$, and $8.5 \times 10^{15} \, 
\mathrm{kg} \, \mathrm{AU}^{-3}$, respectively.  
Predictions about the effects of dark matter on planetary orbits are
potentially observable, as the constraints on Earth and our predictions differ
by less than a factor of two.

Another interesting issue to address is the Pioneer anomaly and the 
possibility that it has arisen due to the dark matter bound to our Solar 
System.  Measurements show that Pioneer 10 and 11 have exhibited 
extra accelerations towards the Sun of $8.09 \pm 0.20 \times 10^{-8} 
\, \mathrm{cm} \, \mathrm{s}^{-2}$ and 
$8.56 \pm 0.15 \times 10^{-8} \, \mathrm{cm} \, \mathrm{s}^{-2}$, respectively
\cite{Anderson:1998}. 
In order for dark matter to have caused this, at least 
$3 \times 10^{-4} M_{\odot}$ 
of dark matter is required within $50 \, \mathrm{AU}$ of the 
Sun \cite{Anderson:1998}. 
The lower bound on the dark matter density capable of causing the anomalies 
is $6.0 \times 10^{18} \, \mathrm{kg} \, \mathrm{AU}^{-3}$ for an inelastic 
scattering of dark matter particles \cite{Adler:2008}. Our results do not
match either of these predictions, as we predict only $\sim 10^{20} \, 
\mathrm{kg}$ of dark matter enclosed within the entire Solar System and a 
dark matter density that never exceeds $2.0 \times 10^{17} \, \mathrm{kg} \,
\mathrm{AU}^{-3}$ anywhere.
We conclude that the Pioneer anamoly cannot be 
caused by dark matter that has been captured by our Solar System. 

Overall, we find that dark matter in our Solar System is far more important
than previously thought.  Due to gravitational three-body interactions
between dark matter particles, the Sun, and the planets, a significant
amount of dark matter winds up gravitationally bound to our Solar System,
resulting in density enhancements between two and five orders of magnitude,
depending on the distance from the Sun.
A future direction for this work is to include back-reaction effects, 
such as subsequent gravitational scatterings of the captured particles.  These
may prove to be important in ejecting a portion of the captured dark matter 
particles, and in reducing the total amount of dark matter that remains 
bound to our Solar System.  More accurate modeling of
our galaxy's dark matter halo, such as the possible inclusion of a 
dark matter disk in the plane of our galaxy \cite{LB}, will also 
alter the net amount of dark matter captured, and is worth further study.
Our results may lead to exciting new directions in 
direct detection experiments and our understanding of dark matter on
the smallest scales.

\begin{acknowledgments}
We thank Romeel Dav\'{e} and Daniel Eisenstein for useful conversations 
concerning this work.  E.R.S. acknowledges NASA grants NNX07AH11G and 
NNX07AC51G as well as NSF AST-0707725 for support.
\end{acknowledgments}



\end{document}